\documentclass[prb,twocolumn,showpacs,amsmath,amssymb]{revtex4}

\usepackage{amsmath}
\usepackage{graphicx}
\usepackage{epsfig}

\newcommand{\be}{\begin{equation}}
\newcommand{\ee}{\end{equation}}
\newcommand{\bma}{\begin{displaymath}}
\newcommand{\ema}{\end{displaymath}}

\begin{document}

\title{Spectral properties of rotating electrons in quantum dots\\
and their relation to quantum Hall liquids}

\author{M. Koskinen$^1$, S.M. Reimann$^2$, J.-P. Nikkarila$^1$ and M. Manninen$^1$}

\affiliation{\sl $^1$NanoScience Center, Department of Physics,
FIN-40014 University of Jyv\"askyl\"a, Finland}

\affiliation{\sl $^2$Mathematical Physics, Lund Institute of Technology,
SE-22100 Lund, Sweden}
 
\date{\today}

\begin{abstract} 
The exact diagonalization technique is used to study 
many-particle properties of interacting  
electrons with spin, confined in a two-dimensional harmonic potential.
The single-particle basis is limited to the lowest Landau level.
The results are analyzed
as a function of the total angular momentum of the system.
Only at angular momenta corresponding to the 
filling factors 1, 1/3, 1/5 etc. the system is fully polarized.
The lowest energy states exhibit spin-waves, domains, and
localization, depending on the angular momentum. 
Vortices exist only at excited polarized states.
The high angular momentum limit shows localization of electrons
and separation of the charge and spin excitations.
\end{abstract}

\maketitle

\section{Introduction}

Semiconductor quantum dots have been 
a rich playground for both experimental and theoretical 
physics\cite{chakraborty1999,reimann2002}. Nearly all methods
of many-particle physics have been used to study the 
ground state properties and excitations in this well-defined
problem of a few electrons trapped in a two-dimensional
harmonic potential. The research has revealed the existence of a 
shell structure, Hund's first rule and many other properties 
related to atomic physics.

In the presence of strong external magnetic fields the quantum dot 
is a finite-size realization of the quantum Hall liquid (QHL).
In fact, it is one of the three geometries (sometimes called
the disc geometry) for performing exact many-particle calculations
for QHLs. The other two geometries are the surface of a sphere, and
a torus~\cite{brey2002,wojs2005,vyborny2006}. 
In a strong magnetic field, the electron system becomes 
polarized. Beyond magnetic field strengths corresponding to 
filling factor $\nu=1$ in QHL (or equivalently, to the maximum density
droplet in a circular quantum dot) the spin does not usually play
a role. Nevertheless, several interesting phenomena, like 
edge reconstruction~\cite{chamon1994,reimann1999}
and the formation of 
vortices~\cite{saarikoski2004,toreblad2004,tavernier2004,saarikoski2004b,toreblad2006} 
as well as magnetic excitations\cite{hawrylak1996,hawrylak1999}
were observed.

The Zeeman effect can be effectively diminished by 
choosing the material parameters such that the effective Land\'e factor
is zero. In so-called bilayer QHLs, a pseudo-spin 
gives the spin-like internal degree of freedom (without the Zeeman effect).
The spin degree of freedom drastically changes the 'simple'
excitations of the QHL. The integer and the
fractional quantum Hall systems at $\nu=1/3$, $1/5$ etc. 
stay polarized\cite{eom2000}
in their ground state and are thus called ferromagnetic,
while at other fractions, like $\nu=2/3$ and $\nu=2/5$, 
the ground state is a spin singlet ($S=0$).
The QHL with spins has been analyzed with numerous theoretical 
methods\cite{morf1986,wu1994,oaknin1996,karlhede1997,falko1999,lejnell1999,walliser2001,smet2001,brey2002,wojs2002,milovanovic2003,doretto2005}.
Many of these studies are aimed at infinite QHLs, using
periodic boundary conditions and (necessarily) restricting the 
numerical diagonalization of the full many-body Hamiltonian  
to a few particles. 

The purpose of this paper is to study systematically
the properties of many-particle spectra of simple 
systems where a small number of electrons is confined
in a two-dimensional harmonic potential. 
We solve the many-particle problem using the configuration
interaction (CI) technique, the only
approximation being that the single-particle basis 
is restricted to the lowest Landau level. 
Since we are mainly interested in large
values of the total angular momentum (corresponding to
small filling factors) this approximation is suitable. 
The results show that many features of the spectrum
are insensitive to the number of particles in the system
and also to the specific form of the 
inter-electron interaction (remember that the Laughlin
Ansatz\cite{laughlin1983} for the fractional Hall effect 
does not contain any information about the Coulomb interaction).

In Section II, we will first describe the theoretical methods used.
In Section III we discuss the results for the ground states and 
low energy excitations.

\section{theoretical methods}

We assume a generic model of interacting electrons
in a two-dimensional harmonic
potential. The Hamiltonian is
\be
H=-\frac{\hbar^2}{2m}\sum_{i}^N \nabla_i^2 
+\sum_i^N \frac{1}{2}m\omega_0^2 r_i^2
+\sum_{i<j}^N \frac{e^2}{4\pi\epsilon_0\vert {\bf r}_i-{\bf r}_j\vert}
\label{hamiltonian}
\ee
where $N$ is the number of particles, $m$ the electron mass, 
${\bf r}=(x,y)$ a two-dimensional position vector, and $\omega_0$ the 
oscillation frequency of the confining potential.
Alternatively, we can use polar coordinates, $x=\Re[r\exp(i\phi)]$,
or complex coordinates $z=x+iy$.
Note that we do not explicitely consider an external magnetic field.
Instead, we solve the many-particle problem for a fixed angular
momentum. In the case of a vanishing Zeeman splitting, the only effect
of the magnetic field is to increase the orbital angular momentum of
the system.

In addition, we also consider a contact interaction between the
electrons. Note that trivially, in this case only electrons with opposite
spins interact: The Pauli exclusion principle
forbids two electrons with the same spin to be at the same point.

The Hamiltonian Eq.~(\ref{hamiltonian}) is solved by using the configuration
interaction (CI) method, with the single-electron basis consisting
of harmonic oscillator states in the lowest Landau level (LLL)
\be
\psi_\ell(r,\phi)=A_\ell r^\ell e^{-m\omega_or^2/2\hbar}e^{i\ell\phi},
\label{spstate}
\ee
where $A_\ell$ is a normalization factor and 
$\ell$ the single-particle angular momentum.
In this basis, the diagonal part of the many-particle Hamiltonian is
constant, $(L+N)\hbar\omega_0$, $L$ being the total angular momentum.
It is then sufficient to diagonalize
the interaction part only. Therefore, the role of the strength 
of the confining potential, $\omega_0$, is only to set the energy
scale. {\it In the LLL the structure of the many-particle 
spectrum and the many-particle wave function are completely 
independent of both $\omega_0$ and the strength of the electron-electron
interaction} ($e^2/4\pi\epsilon_0$).
The restriction of the single-particle space to the LLL yields the 
exact result only in the weak interaction limit. However, independent of
the strength of the interaction, this approximation becomes more and more 
accurate when the total angular momentum of the system
increases\cite{manninen2001}. For the numerical calculation of the 
Coulomb matrix elements, see Stone {\it et al.}~\cite{stone1991}. 

We diagonalize the Hamiltonian matrix for a fixed total angular
momentum $L$ and a fixed $z$-component of the total spin, $S_z$.
We do not fix the total spin, but resolve its value from the 
eigenstates by taking the expectation value of the $\hat S^2$ operator. 
By selecting $S_z=0$ for even number of electrons ($S_z=1/2$ for odd)
the diagonalization of the Hamiltonian gives the energies and
eigenstates for all possible values of the total spin.
The dimension of the Fock space increases fast with $N$ and $L$.
For example, for $N=4$, $L=30$ the matrix dimension is 1234, but
for $N=7$, $L=42$ it is already 43600.
For large systems we make a further reduction of the Fock space
by restricting the maximum single-particle angular momentum of
the space, $\ell\le \ell_{\rm max}$.
The numerical diagonalization was made using the Lanczos method~\cite{arpack}.
It  gives us the ground state and low-lying  
excited states. 

The symmetry of the system requires that
the total electron densities and spin-densities are circularly
symmetric for all the states. In order to study the internal structure,
we thus have to determine correlation functions. Here we use
spin-dependent pair-correlation functions 
\be
g_{\uparrow\sigma}({\bf r},{\bf r}')=\langle \Psi \vert \hat
n_{\uparrow}
({\bf r})\hat n_{\sigma}({\bf r}')\vert \Psi\rangle,
\label{pairc}
\ee
where $\vert \Psi\rangle$ is the 
many-particle state in question, 
$\sigma$ denotes spin up  ($\uparrow $)  or down ($\downarrow $),
and $\hat n_\uparrow$ is the spin-up-density operator. We define
the total pair-correlation as
\be
g({\bf r},{\bf r}')=\frac{1}{2}(g_{\uparrow\uparrow}({\bf r},{\bf r}')+
g_{\uparrow\downarrow}({\bf r},{\bf r}'))~.
\label{pctotal}
\ee
The total angular momentum $L$ can be related to the filling factor
$\nu$ via the relation 
\be
\nu \approx \frac{N(N-1)}{2L}.
\label{ff}
\ee
For small particle numbers, this relation is strictly valid
only for filling factors $\nu=1,~1/3~,1/5,\cdots$, i.e. for those states
of the fractional quantum Hall effect, which can be approximately
described by the Laughlin wave function\cite{laughlin1983}.

Within the LLL, the smallest possible angular momentum is
$L_2=N(N/2-1)/2$ corresponding to a single Slater determinant where the spin-up
and spin-down states are occupied from $\ell=0$ to $\ell=N/2-1$
(for even number of particles). It is natural to associate this state
to filling factor $\nu=2$, while Eq.~(\ref{ff}) would give
$\nu\approx 2(1+1/(N+1))$. Similarly, for fractional filling factors 
2/3 and 2/5 etc., Eq.~(\ref{ff}) gives only an estimate of the 
corresponding angular momentum.

We call the state with filling factor one 
the maximum density droplet (MDD)\cite{macdonald1993}.
In this case, the angular momentum is well defined,
$L_{\rm MDD}=N(N-1)/2$.

At large angular momenta the electrons crystallize in 
a rotating Wigner molecule\cite{reimann2006,nikkarila2006}.
In this case the {\it charge excitations} can be described
by classical vibrations of the molecule,
separated from the {\it spin excitations} of the system.
The equilibrium positions of classical electrons depend on the 
angular velocity $\omega_r$ or angular momentum $L=I\omega_r$
of the Wigner molecule ($I$ is the moment of inertia $I=\sum mr_i^2$),
and they can be solved by minimizing the classical energy
\be 
E_{\rm cl}^0(L)=\frac{1}{2}m\omega_0\sum_i^N r_i^2
+\sum_{i<j}\frac{e^2}{4\pi\epsilon_0\vert{\bf r}_i-{\bf r}_j\vert}
+\frac{L^2}{2m\sum r_i^2}.
\ee
The eigenfrequencies of the vibrations can then be 
solved from the equations of motion of the rotating frame
(by linearizing the equations around the equilibrium positions
of electrons)\cite{nikkarila2006}.
Quantization of the rotational and vibrational modes 
gives an estimate for the energy spectrum
\be
E_{\rm QM}=E_{\rm cl}^0+\sum_k \hbar\omega_k(n_k+\frac{1}{2})+
\hbar\omega_0(n_{\rm 0}+1),
\ee
where $\omega_k$ are all the vibrational frequencies determined in the 
rotating frame and $n_k=0,~1,~2,\cdots$, and the last term
corresponds to the center of mass excitations.

Once localized, the electrons form a system where the charge and 
spin excitations begin to separate. This is most clearly 
seen in one-dimensional systems\cite{kolomeisky1996}.
In the case of quasi-one-dimensional quantum rings 
the whole many-particle spectrum can be quantitatively
described with a model Hamiltonian\cite{koskinen2001,viefers2004}
\be
H_{\rm model}=\frac{\hbar^2 L^2}{2I}+\sum_\alpha
\hbar\omega_\alpha(n_\alpha+\frac{1}{2})
+J\sum_{i,j}\hat S_i\cdot \hat S_j~,
\label{mh}
\ee
which consists of rigid rotations, vibrational modes of the 
localized electrons and of the Heisenberg model for 
the spin configuration.

\section{Results}

\subsection{Oscillating towards ferromagnetism, $2\ge\nu\ge1$}

\begin{figure}[h]
\includegraphics[angle=-90,width=\columnwidth]{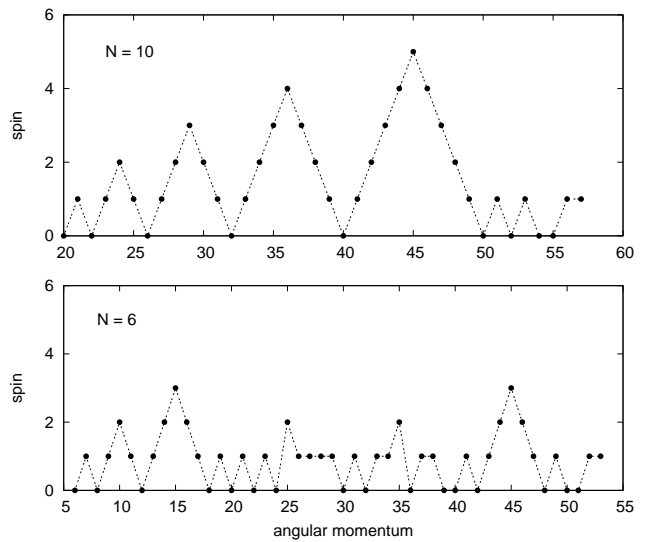}
\caption{Total spin of the lowest energy state as a function of the
angular momentum for $N=10$ and $N=6$ particles. 
}
\label{spin}
\end{figure}

Figure~\ref{spin} shows the total spin for the system
of $N=6$ and $N=10$ electrons as a function of the angular momentum.
The spin increases, oscillating between zero and its 
maximum values, up to $S=N/2$,
which is the spin of the maximum density droplet (MDD).
Note that at this point the electron system is fully
polarized (i.e., ferromagnetic). 
The lowest possible angular momentum for $N=10$ 
(in the LLL) is $L=20$. The corresponding state can be written as
\be
\Psi_{\nu=2}=\prod_{i<j}^{N/2} (z_i-z_j)
\prod_{k<l}^{N/2} (\tilde z_k-\tilde z_l)
e^{-\sum \vert z\vert^2}
\label{mdd2}
\ee
where we denote the coordinates with the spin-down electrons as
$\tilde z$. 
In the occupation number representation we write this 'double
MDD' as 
\bma
\mid \Psi_{(L=20)}\rangle =\left\vert\begin{array}{cccccccccccccc}
\uparrow&\uparrow&\uparrow&\uparrow&\uparrow&0&0&
0&0&0&0&0&0&0\\
\downarrow&\downarrow&\downarrow&\downarrow&\downarrow&0&0&0&0&0&0&0&0&0
\end{array}\right\rangle,
\ema
where the arrows show the occupied states of the LLL
with increasing order of the single-particle angular momentum.
The wave function of the MDD, with $S_z=N/2$, is exactly the
Laughlin state
\be
\Psi_{\rm MDD} =\prod_{i,j}(z_i-z_j) e^{-\sum\vert z\vert^2},
\label{mdd}
\ee
which in the occupation number representation is
\bma
\mid \Psi_{\rm MDD}\rangle =\left\vert\begin{array}{cccccccccccccc}
\uparrow&\uparrow&\uparrow&\uparrow&\uparrow&\uparrow&\uparrow&
\uparrow&\uparrow&\uparrow&0&0&0&0\\
0&0&0&0&0&0&0&0&0&0&0&0&0&0 \end{array}\right\rangle.
\ema
The MDD has a spin degeneracy $2S+1$. For a given $S_z$
the state can be written as
\be
\Psi_{\rm MDD}=\prod_{i<j}^{N/2+S_z} (z_i-z_j)
\prod_{k<l}^{N/2-S_z} (\tilde z_k-\tilde z_l)
\prod_{m,n}(z_m-\tilde z_n)
e^{-\sum \vert z\vert^2}.
\label{mdds}
\ee
It is important to note that for $S_z<N/2$ the above
wave function is a linear combination of several Slater 
determinants. For example, for $S_z=0$, these determinants are
of the type
\bma
\left\vert\begin{array}{cccccccccccccc}
\uparrow&0&0&\uparrow&0&\uparrow&\uparrow&
\uparrow&0&0&0&0&0&0\\
0&\downarrow&\downarrow&0&\downarrow&0&0&0&
\downarrow&\downarrow&0&0&0&0 \end{array}\right\rangle,
\ema
i.e. there is one electron at each angular momentum from 
0 to $N-1$, but its spin can be up ($\uparrow $)  or down ($\downarrow $). 
It is interesting to note that 
in the LLL approximation the exact wave function
of the MDD is the (ferromagnetic)
Laughlin state, although one would expect that much stronger correlations 
could be obtained by also allowing configurations with 
two opposite spins at the same angular momentum.

It is now easy to understand the maxima before the MDD, see Fig.~\ref{spin}.
They consist of two maximum density droplets of different size,
for example 
\bma
\mid \Psi_{(L=24)}\rangle =\left\vert\begin{array}{cccccccccccccc}
\uparrow&\uparrow&\uparrow&\uparrow&\uparrow&\uparrow&\uparrow&
0&0&0&0&0&0&0\\
\downarrow&\downarrow&\downarrow&0&0&0&0&0&0&0&0&0&0&0
\end{array}\right\rangle.
\ema
These states were studied in also in Ref.~\cite{wensauer2003}.
The states in between the maxima are more complicated and 
can no longer be described by a single determinant. 
For example, for $L=40$, the most important configuration
\bma
\left\vert\begin{array}{cccccccccccccc}
\uparrow&0&\uparrow&0&\uparrow&0&\uparrow&0&\uparrow&0&0&
0&0&0\\
\downarrow&0&\downarrow&0&\downarrow&0&\downarrow&0&\downarrow&0&0&0&0&0
\end{array}\right\rangle
\ema
has only 1.5 \% weight in the exact wave function.

Figure \ref{spin} shows that at small angular momenta
(large filling factor) the system has a tendency to ferromagnetism. 
The states between the MDD's can be understood as spin waves
of the ferromagnetic system, as explained in the 
following section. 
The oscillations with increasing angular momentum, until the ferromagnetic 
MDD is reached, see Fig. \ref{spin},
are independent of the particle number.

The results for the LLL in the region $2\ge\nu\ge1$ are
'universal', i.e. {\it independent of the electron-electron
interaction}. We repeated the calculations with 
a contact interaction. The resulting dependence of the 
total spin on the angular momentum was found to be the same as for the Coulomb
interaction. Moreover, the simple states for the spin maxima 
are identical, and those in between are also very similar.
Table \ref{olap1} shows overlaps of the many-particle states
($\vert\langle\Psi_{\rm Coulomb}\vert\Psi_{\rm delta}\rangle\vert^2$)
for different angular momenta. The results demonstrate that for each
angular momentum the lowest energy
state is the same, but the order of the excited states can
depend on the range of the interaction.

\begin{table}[h]
\caption{Overlaps of the wave functions 
calculated with Coulomb and contact interaction
for 10 electrons. $L$ is the angular momentum and $S_i$ the
spin of the state. The subscripts 0 and 1 refer to the lowest energy
state and the first excited state, respectively.
For $L=44$ and 45 the overlaps marked by * are those between the 
first excited state for Coulomb, and the second excited state for
contact interaction. 
}
\begin{tabular}{|l|ll|ll|}
\hline
$L$& $S_0$ & $\vert\langle\Psi_{\rm C}\vert\Psi_{\delta}\rangle_0\vert^2$ &
$S_1$ &$\vert\langle\Psi_{\rm C}\vert\Psi_{\delta}\rangle_1\vert^2$ \\
\hline
40 & 0 & 0.979 & 1& .979 \\
41 & 1& 0.985 & 2& .985 \\
42 & 2& 0.991 &3&  .991 \\
43 & 3& 0.996 & 4& .997 \\
44 & 4&0.999 & 2& .785$^*$ \\
45 & 5&1 & 3&.977$^*$ \\
\hline
\end{tabular}
\label{olap1}
\end{table}

For the MDD the overlap equals one, since in both cases 
the state is the simple Laughlin state of Eq.~(\ref{mdds}) (consisting
for $N=10$ electrons of 256 Slater determinants with the same weight).
To appreciate the good overlap between the more complicated states,
one should note that, for example, for $L=40$ 
the 1000 most important Slater
determinants only contribute 97.0 \% of the total wave function.

\subsection{Spin waves as excitations of the ferromagnetic state}

The completely polarized MDD is the ferromagnetic  
integer quantum Hall state~\cite{ho1994,fertig1994,eom2000}. 
When the angular momentum is increased, the polarization
decreases linearly to zero 
at angular momentum $L=L_{\rm MDD}+N/2$. 
At larger angular momenta, the total spin of the lowest energy state
is small, usually 0 or 1, until one approaches the filling
factor $\nu=1/3$ where the system again becomes polarized,
as seen in Fig. \ref{spin} for six electrons ($\nu=1$ corresponds 
to $L=15$ and $\nu=1/3$ to $L=45$).

\begin{figure}[h]
\includegraphics[angle=-90,width=\columnwidth]{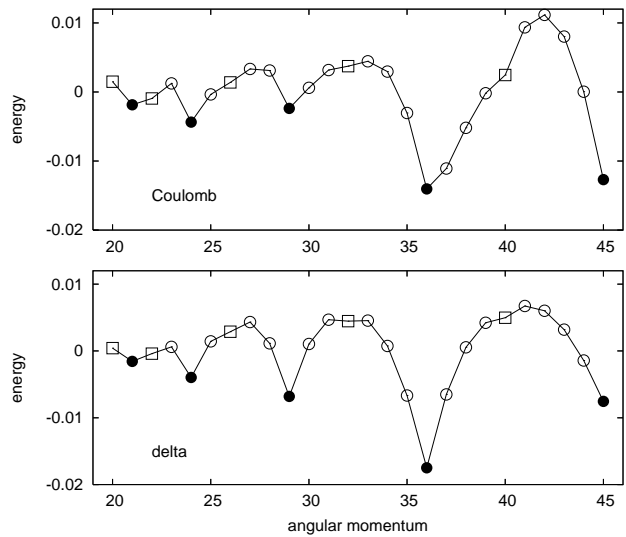}
\caption{Total energy as a function of the angular momentum 
for 10 electrons interacting with Coulomb interaction (upper panel)
or delta function interaction (lower panel). Black dots indicate the
states with local spin maxima (see Fig. \ref{spin}) and the squares
the $S=0$ states. The energy is obtained by subtracting 
a third order polynomial fit from the total energy.
}
\label{edif}
\end{figure}

Figure \ref{edif} shows the oscillations in the total energy 
for 10 electrons as a function of
the angular momentum. The curves are obtained by subtracting a third-order 
polynomial fit from the total energy. 
The oscillating part of the total energy, here in the interval $L=20$ 
to $L=45$, is similar for Coulomb and contact interactions.
The downwards cusps correspond to the spin maxima.
As explained above, these states consist of two maximum
density droplets with different size for spin-up and spin-down 
electrons.

\begin{figure}[h]
\includegraphics[angle=-90,width=\columnwidth]{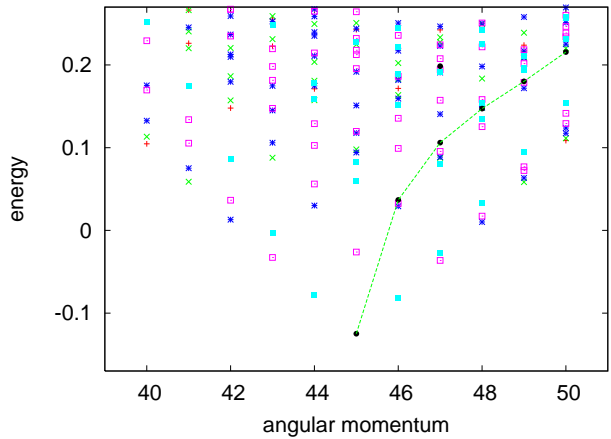}
\caption{Energy spectrum around the maximum density droplet of
10 electrons. Different symbols denote different spin:
$S=5$, black dots; $S=4$, blue filled squares; $S=3$, red open squares;
$S=2$, blue stars; $S=1$, green $\times$; $S=0$, red $+$.
The lowest energies for the fully polarized states are connected with
a line.}
\label{n10l45s}
\end{figure}

To understand the nature of the other states, 
we first study the excitation spectrum at the vicinity of 
the MDD. Figure \ref{n10l45s} shows the energy spectrum for 10
particles from $L=40$ to $L=50$.
To illustrate the symmetry of the spectrum around the 
MDD ($L=45$), we subtract a linear function 
from the total energy: Figure~\ref{n10l45s} shows
$\Delta E=E-\hbar\omega_0(L+1)-19.86+0.162 L$,
where the last two terms are obtained by fitting a linear 
function of the interaction energy (between $L=40$ and
$L=50$). The spectrum is shown for the Coulomb interaction. 
In the case of the contact interaction, 
the energy differences between
the different spin states disappear after the MDD.
The reason is that the MDD wave function, Eq.~(\ref{mdd}),
has zero amplitude whenever two electrons are at the same 
position and, consequently, the interaction energy of the 
contact interaction is zero. 
Beyond the MDD, the wave function can be constructed
by multiplying the MDD state with a symmetric homogeneous
polynomial\cite{ruuska2005}.

Figure \ref{n10l45s} has a remarkable similarity with results in the study by 
W\'ojs and Quinn\cite{wojs2002,wojs2005}, who considered a small number of 
electrons on the surface of a sphere.
A similar spectrum is also obtained when using the geometry 
of a torus\cite{vyborny2006}. The structure of the spectrum seems
then to be independent of the boundary conditions, though
in each case the meaning of the angular momentum is
different.

The polarized state is the lowest-energy state 
only at the MDD (and at the higher Laughlin states).
For other angular momenta, the spin is reduced. 

The spin of the ferromagnetic MDD 
can be reduced by spin-wave excitations.
The operator that excites spin waves
can be written as~\cite{oaknin1996}
\be
\Sigma_1^+=\sum_{\ell=0}^N \sqrt{\ell+1}~c_{\ell+1\downarrow}^+c_{\ell\uparrow}
\label{spinv}
\ee
where $c$ is the normal fermion annihilation operator.
Note that operating with $\Sigma_1^+$ to the MDD the angular momentum is
increased by one and the total spin is decreased by one.
Similarly, we can form spin waves which lower the angular momentum:
\be
\Sigma_1^-=\sum_{\ell=1}^N \sqrt{\ell}~c_{\ell-1\downarrow}^+c_{\ell\uparrow}
\label{spinvd}.
\ee
Moreover, assuming the spin waves as independent excitations, as 
in the theory of ferromagnetism~\cite{ashcroft1976}, we can create two
or more spin waves by operating successively with the above operators,
for example, the state with two spin waves and angular momentum
$L=L_{\rm MDD}+2$ is (without normalization)
\begin{eqnarray}
\vert \Psi_{2{\rm SW}}\rangle &=&
\sum_{\ell_1<\ell_2}^N 
\sqrt{(\ell_1+1)(\ell_2+1)}\\
& &c_{\ell_1+1\downarrow}^+c_{\ell_1\uparrow}\nonumber
c_{\ell_2+1\downarrow}^+c_{\ell_2\uparrow}\vert \psi_{\rm MDD} \rangle.
\label{spinv2}
\end{eqnarray}

\begin{table}
\caption{Overlap between the exact result and the
spin waves states. $N$, $L$, and $S$ are the number of electrons,
angular momentum and total spin, respectively. $n$ is the number of
spin waves.  
}
\begin{tabular}{llllr}
\hline
$N$ & $L$& $S$& n& $\vert\langle\Psi\vert\Psi_{n{\rm SW}}\rangle\vert^2$\\
\hline
10 & 40 & 0 & 5 &  .296   \\
10 & 41 & 1 & 4 & .647 \\
10 & 42 & 2 & 3 & .844 \\
10 & 43 & 3 & 2 & .941 \\
10 & 44 & 4 & 1  & .986 \\
10 & 45 & 5 & 0 & 1 (MDD)\\
10 & 46 & 4 & 1 & .972  \\
10 & 47 & 3 & 2 & .913  \\
10 & 48 & 2 & 3 & .803  \\
10 & 49 & 1 & 4 & .607  \\    
10 & 50 & 0 & 5  & .262     \\
18 & 153 & 9 & 0 & 1 (MDD) \\
18 & 154 & 8 & 1 & .990 \\
18 & 155 & 7 & 2 & .971  \\
18 & 156 & 6 & 3 & .957  \\
\hline
\end{tabular}
\label{olap3}
\end{table}

Table \ref{olap3} shows overlaps between the 
exact result and that of the spin wave approximation.
For the single spin wave the results agree with those of 
Oaknin {\it et al.}\cite{oaknin1996}. For $N=10$ electrons, 
the overlap decreases with the number of the
spin waves, but is still more than 60 \% for four spin
waves. However, for the singlet state, which requires 
five spin waves, the overlap is less than 30 \%. 
Clearly, this singlet state has a different character.
This state ($L=40$ for $N=10$) also shows a small kink in the
total energy as a function of the angular momentum
(see Fig. \ref{edif} above). We will return to this state in the 
next section.
Oaknin {\it et al.}\cite{oaknin1996} have shown that 
by projecting the simple spin wave state, Eq.~(\ref{spinv}), 
to a state which is orthogonal
to the center-of-mass excitations, the overlaps become 
even better.

\begin{figure}[h]
\includegraphics[angle=-90,width=\columnwidth]{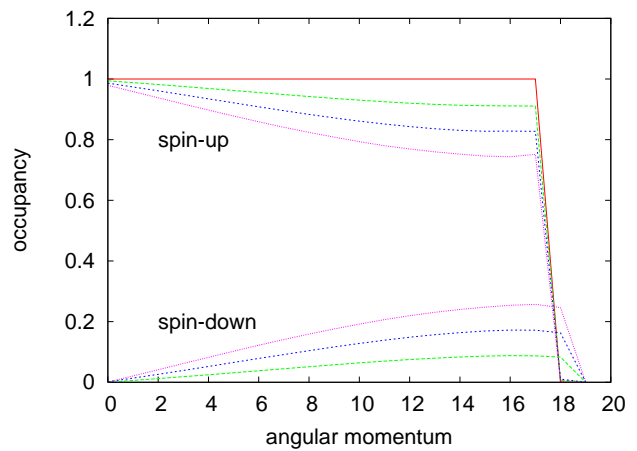}     
\caption{Occupation of the single-particle 
levels of the yrast state at different angular momenta
for 18 electrons.
Red line: $L=153$ (MDD), green line: $L=154$ 
blue line: $L=155$, pink line: $L=156$.
}
\label{occ18}
\end{figure}

Table \ref{olap3} shows that when the number of electrons increases,
the approximation of independent spin waves gets more accurate.
For $N=18$ electrons, even the three spin waves describe the exact
result with 96 \% accuracy. 
Figure \ref{occ18} shows the single-particle occupancy
of the exact states corresponding to 1, 2 and 3 spin waves
for 18 electrons. Clearly, the occupancy of the minority
spin states (spin-down) increases linearly as a function of the 
single-particle angular momentum (for small angular momenta)
in agreement with the spin wave suggestion, Eq. (\ref{spinv}).
Moreover, the increase of the spin waves increases the occupancy
linearly. These observations are in qualitative agreement with
the theory of Doretto {\it et al.}\cite{doretto2005},
which shows that the magnetic excitons have bosonic nature.

\begin{figure}[h]
\includegraphics[angle=-90,width=\columnwidth]{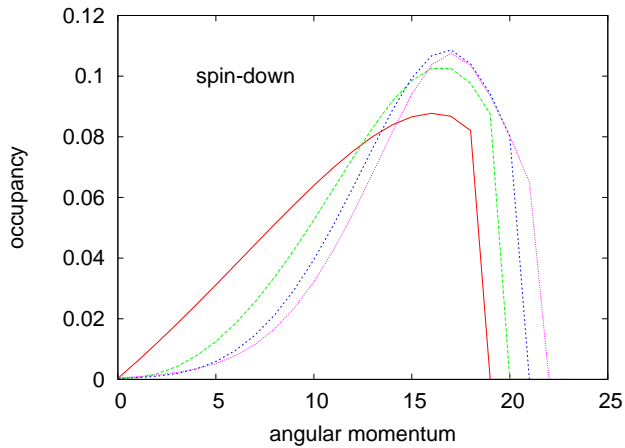}
\caption{Occupation of the single-particle 
levels of the lowest $S=8$ state at different angular momenta
for 18 electrons.
Red line: $L=154$ (MDD), green line: $L=155$ 
blue line: $L=156$, pink line: $L=157$. 
(The red line is the same as the lower green line in Fig. \ref{occ18}.
Note the different scale).
}
\label{occ18b}
\end{figure}  

Increasing the angular momentum from the MDD, the energy spectrum
shows a large energy gap between low energy states and the rest of the
states (including the fully polarized state, see Fig. \ref{n10l45s}).
We demonstrated previously that the lowest of these states consist
of spin-waves. The highest state below the energy gap has
spin $S=N/2-1$ in the whole region from $L=L_{\rm MDD}+1$ up to 
$L=L_{\rm MDD}+N/2$ (i.e. from $L=46$ to $L=50$ for $N=10$). 
Figure \ref{occ18b}
shows that the minority spin concentrates closer and closer to the 
surface when the angular momentum increases. 
The total spin density $n_\uparrow-n_\downarrow$ still remains
positive at the surface. This  suggests  that the
spins along the surface are slightly tilted\cite{heinonen1999}. 
These states are often referred to as having non-collinear (or canted)
spins\cite{heinonen1999,freeman2002}.
We should note that in our computations we have chosen $S^2$ and $S_z$ 
as good quantum numbers. Consequently, we do not know the 
values of $S_x$ and $S_y$.

\subsection{Domain walls}

Large ferromagnetic systems may form domains. In normal
magnetic materials the domains are caused by the long-range dipole 
interaction and exist already in the ground state. However,
even without the dipolar interaction, domains can form as low energy 
excitations. This seems to be the case in  
quantum Hall 
liquids\cite{falko1999,eom2000,smet2001,abolfath2002,chalker2002,brey2002,rezayi2003}.

\begin{figure}[h]
\includegraphics[angle=-90,width=\columnwidth]{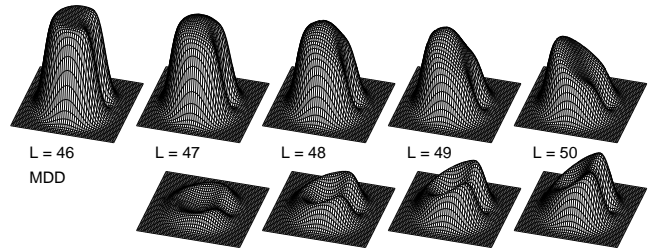}
\caption{Pair correlation functions beyond the MDD for $N=10$ electrons.
The upper panel shows the up-down correlation $g_{\uparrow \downarrow}$, 
and the lower panel the up-up correlation, 
$g_{\uparrow \uparrow}$. 
The number of spin-down electrons increases from 
one to five, and the number of spin-up electrons decreases from 9 to 5
when going from left to right.
}
\label{domains1}
\end{figure}

\begin{figure}[h]
\includegraphics[angle=-90,width=\columnwidth]{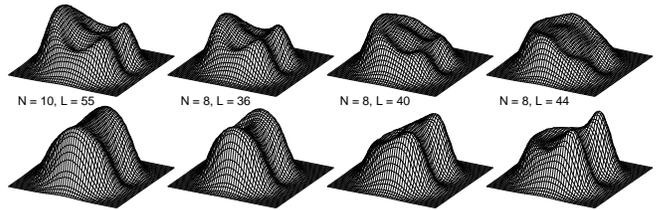}
\caption{Pair correlations for spin-singlet states after the MDD.
The upper row shows the up-down correlation, $g_{\uparrow, \downarrow }$, 
and the lower row the up-up correlation,  $g_{\uparrow, \uparrow }$. 
The number of electrons $N$ and the angular
momentum $L$ are indicated. Note that $N=10$, $L=55$ corresponds to the 
state $N=8$, $L=36$.}
\label{domains2}
\end{figure}

In finite small systems, the existence of domains is not as obvious,
due to the small number of electrons and the finite thickness of the 
domain walls. Nevertheless, our results show a clear separation of up and 
down spins at certain angular momenta. Figures \ref{domains1} 
and \ref{domains2} show
a set of pair correlations showing the tendency to separate the
spin-up and spin-down electrons. The gradual development to a 
'domain wall state' is indicated in Fig.~\ref{domains1}, which shows the
correlations for a dot with $N=10$ electrons from angular 
momenta $L_{\rm MDD}$ to $L_{\rm MDD}+N/2$. 
The largest of these  angular momenta ($L=50$) shows clearly, that
the up and down electrons prefer to be at the opposite sides of the 
dot. (If the reference electron, say spin-up, was
at the center of the dot, the other spin-up electrons would 
stay closer to the origin, while the spin-down electrons are pushed 
further out).

The structure of two domains for $L=L_{\rm MDD}+N/2$ is a general
feature and does not depend on the number of electrons. 
Calculations were performed for 
$N=6, 8, 10,$ and 12 electrons. In all cases, the
development of the pair correlation functions appear to 
be similar to that shown in Fig. \ref{domains1} above. 

Increasing the angular momentum further results in more complicated
domain structures. For angular momentum $L=L_{\rm MDD}+N$, four domains
seem to form as seen in Fig. \ref{domains2} for
$N=8$, $L=36$ and $N=10$, $L=55$. 
For even larger angular momenta it is not possible to resolve
possible domain structures at these small particle numbers.
At angular momentum $L=44$, corresponding to the filling factor
$\nu=2/3$, the pair correlation function can not be as easily interpreted.

It is important to note that our computations do not contain the
dipole-dipole interaction of the magnetic moments of electrons. 
The origin of the domain formation is thus different from that in 
normal ferromagnets. An infinite system would not have domains in the
ground state. The fact that they appear in the lowest energy state 
at a fixed angular momentum does not imply that they represent a ground
state.
The antisymmetry of the total wave function 
restricts the allowed quantum states of a given total angular
momentum. For example, the polarized state has lowest 
energy only for a few angular momenta, while others must have some
other internal structure. This becomes clear in the limit of
localized electrons, to be discussed later.

\subsection{Vortices and edge reconstruction}

\begin{figure}[h]
\includegraphics[angle=-90,width=\columnwidth]{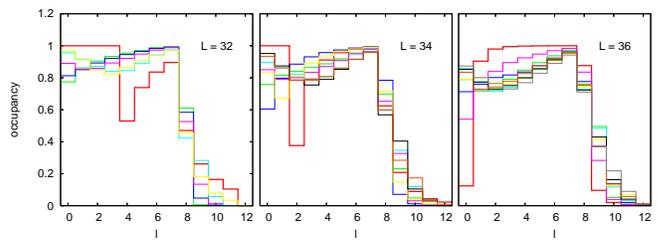}
\caption{Single-particle occupancy of different energy states 
for $N=8$, $L=32$, 34, and 36. The occupancies of the 
lowest fully polarized state $S=4$ are shown by thick red lines,
the states below the first polarized state are shown by thin lines.
The polarized results show a clear dip at a given single-particle
state as a signature of a single vortex, which reaches origin 
at $L=36$.
}
\label{vortex1}
\end{figure}

In a strong magnetic field, the Zeeman effect polarizes the electrons
and the system usually stays polarized beyond the maximum density
droplet. The low-energy excitations of the MDD are then 
characterized by vortices~\cite{saarikoski2004,toreblad2004,tavernier2004,saarikoski2004b} or edge reconstruction~\cite{macdonald1993,reimann1999}.
As discussed earlier by Yang {\it et al.}~\cite{yang1993},
when the angular momentum increases from that of the 
MDD, {\it holes} will be formed in the otherwise filled
Fermi sea below the single-particle angular momentum $\ell=N-1$.
Close to the MDD, the most important configuration of the many-particle
state will have only one hole. For example at $L=32$, $N=8$ we have
\bma
\left\vert\begin{array}{cccccccccccc}
\uparrow&\uparrow&\uparrow&\uparrow&0&\uparrow&\uparrow&\uparrow&
\uparrow&0&0&0\\
0&0&0&0&0&0&0&0&0&0&0&0 \end{array}\right\rangle.
\ema
Increasing the angular momentum, the hole stepwise decreases its 
(single-particle) angular momentum, ``moving to the left'' in the Fock state 
until it reaches $\ell=0$.
In the full CI calculation for polarized electrons, there are
many other possible configurations with the same angular momentum, but 
this single-hole configuration has the largest weight. This is clearly seen
in Fig. \ref{vortex1}, where the thick red line shows the 
result for polarized electrons ($S=4$).

The formation of a single hole in the MDD~\cite{yang1993} 
can be understood in two ways. We can consider it as a single-particle
excitation, where an electron is excited from
$\ell=4$ to $\ell=8$. Alternatively, we can view it as a collective
excitation, where four electrons from $\ell=4\cdots 7$ are each
excited to the next angular momentum state. The latter interpretation
is consistent with the vortex picture in boson 
systems\cite{toreblad2004} where the Bose condensate
corresponds to the MDD of spinless fermions.

The holes in the electron sea of the MDD can be identified with 
vortices~\cite{manninen2005}. The electron density has a minimum at the 
site of the vortices,
the wave function phase changes by $2\pi$ in going around  
the vortex center~\cite{toreblad2004}  and the electron current 
circulates around the vortex~\cite{saarikoski2004}.

We now return to the case without Zeeman splitting. The total
spin of the lowest-energy states beyond the MDD is then
generally small, and the fully polarized state is 
a rather high-lying excited state. Figure \ref{vortex1} shows 
the single-particle occupancies for all the energy
states below the polarized states, at angular momenta $L=32, 34$ and $36$.
A typical dip in the 
occupancy at one single-particle angular momentum is seen only 
in the fully polarized state. The low-spin states
show a more uniform reduction of the occupancy and softening of 
the surface. Vortices seem to be low-energy excitations only
for the polarized electron gas.

\begin{figure}[h]
\includegraphics[angle=-90,width=\columnwidth]{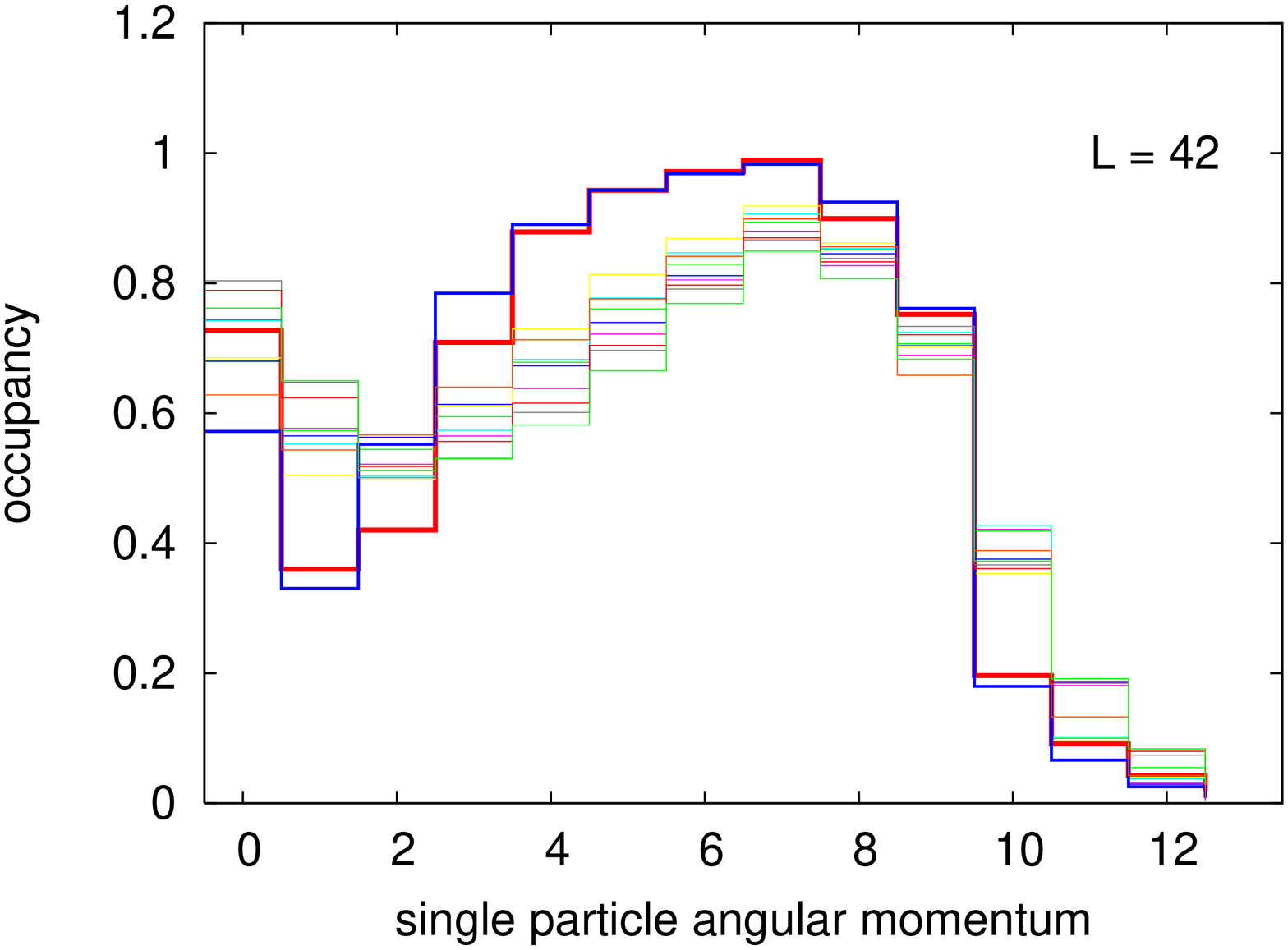}
\includegraphics[angle=-90,width=\columnwidth]{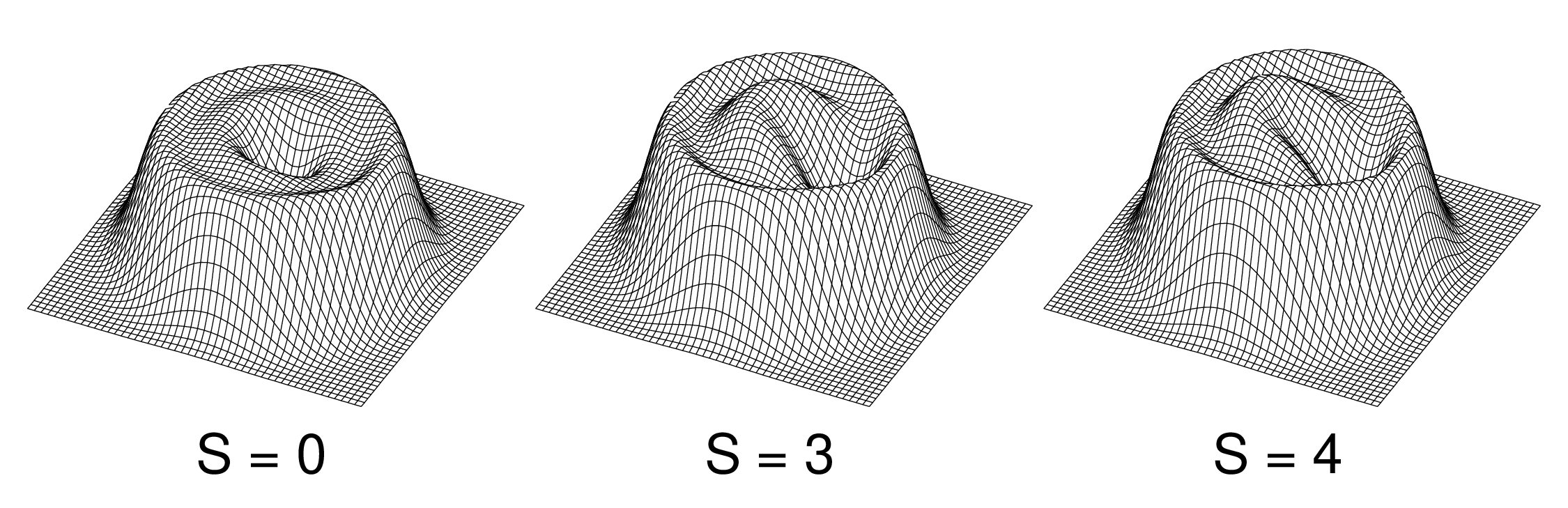}
\caption{Upper panel: Occupancy of the single-particle angular momenta
of the lowest state for $N=8$, $L=42$ (red line), and for the lowest 
$S=3$ (blue line) and $S=4$ (green line) states.
The lower panel shows the hole-hole correlation functions 
for the same states. 
}
\label{vortex2}
\end{figure}  

The most clear two-vortex state in the case of eight electrons is 
at angular momentum $L=42$ (for polarized
electrons)\cite{toreblad2004}. In this case, the most important 
configuration is
\bma
\left\vert\begin{array}{cccccccccccc}
\uparrow&0&0&\uparrow&\uparrow&\uparrow&\uparrow&\uparrow&\uparrow&
\uparrow&0&0\\
0&0&0&0&0&0&0&0&0&0&0&0 \end{array}\right\rangle.
\ema
More generally, for $n$ vortices the most important configuration
is the one where the $n$ holes are next to
each other\cite{manninen2005}. Figure \ref{vortex2} shows the
single-particle occupancies of the exact diagonalization for
$N=8$, $L=42$ for all the states up to the fully polarized state.
Again, we see that only in the polarized state,
shown by the thick red line,
there clearly are two adjacent states ($\ell=1$ and $\ell=2$) where
the occupancy is small compared to the other states. 
The only other state which has a rather similar occupancy distribution,
is the one with total spin $S=3$, i.e. the almost fully polarized state,
shown by the thick blue line. 
The other spin states show only a weak minimum at the region
of $\ell=1$ or 2, but larger occupancies from $\ell=10$ up, i.e.
at the surface of the electron cloud. In fact, the general feature of
the occupancy distribution for all the states is, that there is a
maximum at $\ell=0$ and a broad maximum at larger $\ell$, the 
center being at about $\ell=7$. This is the first indication
that at high (total) angular momenta the electrons begin to localize.
For $N=8$ electrons, eventually one electron will localize at the
center, while the other seven form a ring around it.

The excited states with $S=4$ and $S=3$ show occupancy distributions
consistent with existence of two vortices\cite{toreblad2004}. 
Other signatures of vortices are
the phase of the wave function, and the pair correlation functions.
The phase is not easy to compute in our formalism. Furthermore, as
it is a function of $2N$ coordinates its
interpretation is not straightforward\cite{toreblad2004}.
Manninen {\it et al.}\cite{manninen2005} have shown that by
interpreting the vortices as holes, their localization can
be seen clearly in the hole-hole correlation function. 
Figure \ref{vortex2} shows the hole-hole correlations for
the lowest energy state $S=0$ and for the $S=3$ and $S=4$ states.
The correlations for the high spin states appear very similar.
The reference vortex (or hole) is seen as a deep minimum. At the
opposite side, there is a maximum corresponding to the second vortex.
The ring around is caused by the holes outside the electron
distribution. (In reality, the hole density is constant
outside the electron distribution, but since here 
the available single-particle space is limited, 
the hole density also goes to zero).
The hole-hole correlation function for the $S=0$ state does not even
show a clear minimum at the site of the reference point (it is filled by 
the hole with opposite spin). The minimum seen in the 
case of $S=0$ is at the
center of the dot and is caused by the fact that the electron density
there is large and, consequently, the hole density is small.

For large electron numbers ($N\gtrsim 20$) the vortices 
of the polarized system can be seen 
clearly in the many-particle spectrum. The localization 
of vortices causes
periodic oscillations to the spectrum\cite{manninen2005} 
in a similar fashion as the localization of the electrons
when their number is small\cite{reimann2006}.  
Unfortunately, we are not yet able to do accurate
computations for such large systems,  when including  
the spin degree of freedom.
However, the above results for $N=8$ electrons strongly suggest
that while the vortices are low-energy excitations for the
polarized electrons, they are not the lowest excitations for
electrons with spin.

In the case of polarized electrons, the MDD breaks up by 
separating a ring of electrons~\cite{chamon1994,reimann1999}.
In the single-particle occupancy, the formation of this
so-called Chamon-Wen edge~\cite{chamon1994} is seen as a minimum in a similar way 
as the formation of vortices. Indeed, these two phenomena 
are intimately related. The minimum between the separated electron
ring and the rest of the MDD is actually caused by one or two 
vortices~\cite{toreblad2006,reimann2006b}. 
In the circularly symmetric electron density, the ring
of vortices is seen as a minimum. In the
vortex-vortex (hole-hole) correlation, however, the vortices are localized 
along this ring. Again, at present we
can not perform computations for large enough electron numbers to
study the possibility of Chamon-Wen edge reconstruction in the low-spin states.
However, due its close relation to the vortex formation,
we can safely predict that it will only be formed in the
excited states with high total spin.

\subsection{Spectra between $\nu=1$ and $\nu=1/3$}

\begin{figure}[h]
\includegraphics[angle=-90,width=\columnwidth]{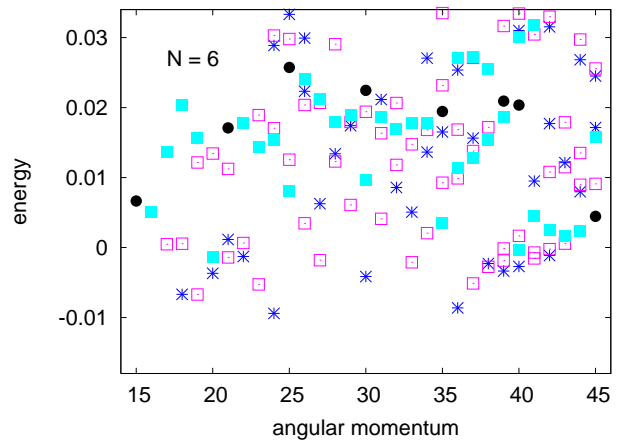}
\caption{Spectrum for $N=6$ electrons for filling factors between 
$\nu=1$ and $\nu = 1/3$, as a function of angular momentum. 
Blue crosses: $S=0$; blue squares: $S=1$, red open squares: $S=2$,
black bullets: $S=3$. A third order polynomial fitted to the lowest
energies has been subtracted from the total energy.
}
\label{spec6}
\end{figure}

\begin{figure}[h]
\includegraphics[angle=-90,width=\columnwidth]{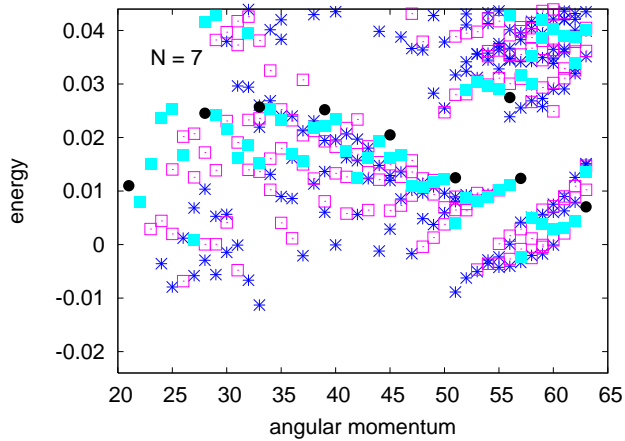}
\caption{Spectrum for 7 electrons $\nu=1 \cdots 1/3$.
Blue crosses: $S=1/2$; blue squares: $S=3/2$; red open squares: $S=5/2$;
black dots: $S=7/2$. 
A third order polynomial fitted to the lowest
energies has been subtracted from the total energy.
}
\label{spec7}
\end{figure}  

\begin{figure}[h]
\includegraphics[angle=-90,width=\columnwidth]{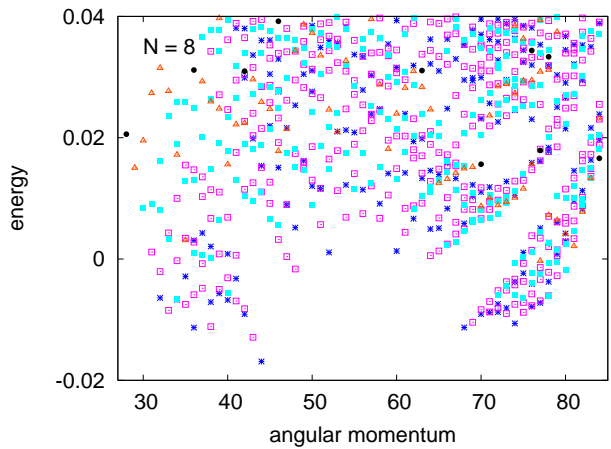}
\caption{Spectrum for 8 electrons $\nu=1 \cdots 1/3$.
Blue crosses: $S=0$; blue squares: $S=1$; red open squares: $S=2$;
red triangles: $S=3$; black dots: $S=4$. 
A third order polynomial fitted to the lowest
energies has been subtracted from the total energy.
}
\label{spec8}
\end{figure}  

The many-particle excitation spectrum shows a remarkable reflection
symmetry around the angular momentum $L=N(N-1)$, corresponding to
filling factor $\nu=1/2$. The composite fermion picture of 
Jain~\cite{jain1989,jain1998} reproduces this symmetry\cite{kamilla1995} 
in a very natural way. Subtracting a smooth 
function of angular momentum, $F(L)$, from the original spectrum
removes the downward slope of the
interaction energy, so that the small energy differences between the 
excited states can be more clearly illustrated, 
and the symmetry around the $\nu =1/2$ states is mapped out in a 
very clear way. 
Figures  \ref{spec6}, \ref{spec7}, and \ref{spec8} show the spectra
for 6, 7, and 8 electrons. In these cases, the function $F(L)$ 
is a third order polynomial fitted in each case to the lowest
energy states in the interval shown.
At filling factor 1/2, the lowest state has total spin 
0, 1/2 and 1, for $N=6$, 7, and 8, respectively.
At $L=N(N-1)/2$ and $L=3N(N-1)/2$ ($\nu=1$ and $\nu=1/3$) 
the lowest state is ferromagnetic
(has maximum spin). 
There are two other energetically favorable
states, at $L=N(N-1)/2+N^2/4$ and $L=3N(N-1)/2-N^2/4$. Our
interpretation is that the former of these
corresponds to the filling factor $\nu=2/3$ and the latter to $\nu=2/5$
(note that Eq.~(\ref{ff}) does not give exactly the above angular
momenta for these special filling factors). The structure of the 
spectra around these special points is independent of the number 
of electrons (for an odd number of electrons, the lowest spin is 
naturally $S=1/2$).

The maximum spin state, $S=N/2$, is the lowest energy state only at
angular momenta corresponding to the filling factors $\nu=1$ (MDD) and 
$\nu=1/3$. It is interesting to observe that while the MDD has a large 
energy gap to the first excited state, the filling factor $\nu=1/3$
does not. In fact, the energy gaps of the special singlet states at
filling factors $\nu=2/3$ and $2/5$ have much larger energy gaps
between the ground state and the lowest excited state.
The large excitation gaps at these 
filling factors are well-known, see for example Ref.~\cite{vyborny2006}.

\subsection{Halperin-Haldane wave functions}

The clear local minimum in the energy at angular momenta 
$L=N(N-1)/2+N^2/4$ and $L=3N(N-1)/2-N^2/4$ suggests that there
is a special way to build correlations in the wave function.
The Ansatz wave function for the fractional quantum Hall
effect\cite{laughlin1983}, Eq.~(\ref{mdds}), can be extended for
some of the non-simple fractions as\cite{halperin1984,balatsky1991}
\be
\Psi_{\rm HH}=\prod_{i<j}^{N/2} (z_i-z_j)^q
\prod_{k<l}^{N/2} (\tilde z_k-\tilde z_l)^q
\prod_{m,n}^{N/2}(z_m-\tilde z_n)^p
e^{\sum \vert z\vert^2}
\label{hh}
\ee
where $q$ is an odd integer and $p$ a positive
integer which can be even or odd. The angular momenta 
$L=N(N-1)/2+N^2/4$ and $L=3N(N-1)/2-N^2/4$ agree with the above
wave function with $q=1$, $p=2$ and $q=3$, $p=2$, respectively. 
Table \ref{olap4} shows the calculated
overlaps. For the $L=N(N-1)/2+N^2/4$ state the overlap decreases rapidly 
with increasing $N$. Clearly, the exact quantum state can not
be described by Eq.~(\ref{hh}) for $\nu=2/3$. 
In fact, in this case the wave function Eq.~(\ref{hh}) is a mixture 
of the $S=0$ and $S=2$ states. 
For $L=3N(N-1)/2-N^2/4$ the agreement is much better, indicating that 
the above wave function is a good approximation
to the ground state of $\nu=2/5$. Also, it now has $S=0$.

\begin{table}
\caption{Overlap between the exact result and the
Halperin-Haldane model, $\vert\langle \Psi_{\rm
  HH}\vert\Psi\rangle\vert^2$,
for different filling factors $\nu$ and electron numbers $N$
}
\begin{tabular}{lllll}
\hline
$N$ $\backslash$ $\nu$ & $1$& $\frac{2}{3}$& $\frac{2}{5}$& $\frac{1}{3}$\\
\hline
2 & 1 & 1 & 1 & 1 \\
3 & 1 & .843 & .920 & .982 \\
4 & 1 & .636  & .931  & .958 \\
5 & 1 & .363  & .911  & .970 \\    
6 & 1 & .162  & .909  & .980 \\
\hline
\end{tabular}
\label{olap4}
\end{table}

It should be noted that for smaller filling factors than 1/3
(or 1/5) the above simple analytic Ansatz does not accurately
describe the true electron state. In particular, it fails to describe
the electron localization. In this region, other trial wave functions, 
like for example 
the ones suggested by Yannouleas and Landman~\cite{yannouleas2002},
seem to be a better alternative. 

\subsection{Symmetry of the spectrum around $\nu=1/3$}

\begin{figure}[h]
\includegraphics[angle=-90,width=\columnwidth]{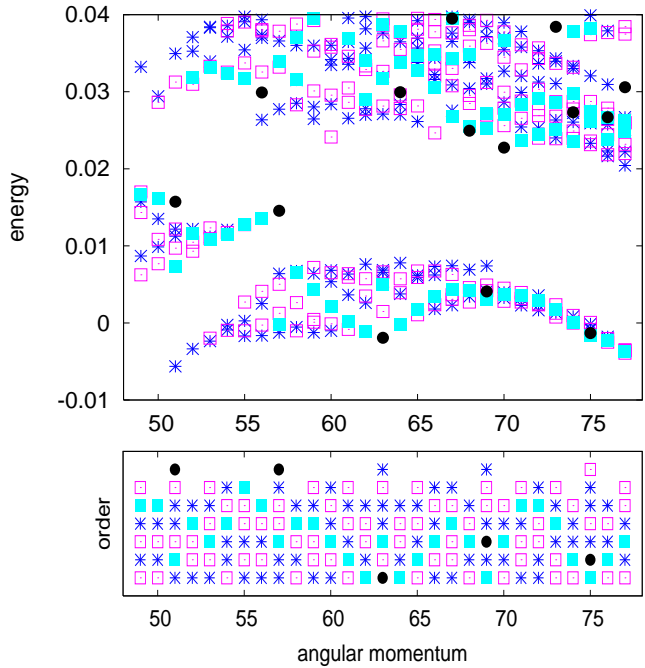}
\caption{Spectrum for $N=7$ electrons around angular momentum $L=63$ 
($\nu = 1/3$). 
Blue crosses: $S=1/2$; blue squares: $S=3/2$; red open squares: $S=5/2$;
black bullets: $S=7/2$. 
The upper panel shows the energy spectrum with a smooth part subtracted.
The lower panel shows the order of different spin states in the 
low-energy band, below the large energy gap in the spectrum. 
Note the symmetry around $L=63$.}
\label{spec7d}
\end{figure}  

In figure \ref{n10l45s} above, we showed that after subtracting the smooth
part, the resulting energy spectrum is symmetric around the MDD ($\nu=1$). 
It is thus interesting to see if similar
symmetry exists also around the angular momentum $L=3N(N-1)/2$, 
corresponding to $\nu=1/3$.
Figure \ref{spec7d} shows the excitation spectrum for 7 electrons, 
now plotted such that the center is around $L=63$. Again, a  
smooth function (in this case, a second order polynomial
fitted to the lowest
energies in the angular momentum interval shown) is subtracted from
the original spectrum. The figure shows that in the immediate
vicinity, from $L=58$ to $L=68$ the spectrum shows a qualitative 
mirror symmetry for the states in the low-energy band, below the large
energy gap. The lower panel of Fig. \ref{spec7d} shows the order of different
spin states for the low energy band. We can see that, in addition to
the symmetry near $L=63$, there is a repeating period of six
in the appearance of different spin states in the whole
range show (ignoring their order). This is due to the emergence of
the localization of
the electrons in a six-fold ring with one electron at the center,
as briefly discussed in the next section. 

\subsection{Localization of electrons at high angular momentum, $\nu
  \ll 1/3$}

Increasing the angular momentum, the electron cloud in the harmonic
confinement expands and the electrons begin to localize. For 
spinless electrons, it is possible to perform accurate CI
calculations for much higher angular momenta than if spin 
is included.
For small electron numbers,
the spectrum shows a  characteristic periodicity as a function of
the angular momentum~\cite{girvin1983,wojs1997,manninen2001,reimann2002,reimann2006,nikkarila2006}.
In these studies the number of electrons has been so small, that 
filling factor down to $\nu=1/9$ have been reached. The results are then
consistent with the expectation that
in quantum Hall liquids the crystallization is 
expected~\cite{lam1984} to happen at filling factors smaller than $\nu=1/7$.
The periodicity in the spectrum as a function of angular momentum is
determined  by the symmetry group of the 
Wigner molecule: Up to five electrons, the 
localized electrons form a single ring, the period being equal to the
number of electrons. From $N=6$ to 8 electrons, one of the 
electrons localizes at the center of the parabolic trap, 
while the rest form a single ring around it\cite{bolton1993,bedanov1994}. 
The length of the period of the
oscillations in these cases is then $N-1$.
From figures \ref{spec7} and \ref{spec8} we can see clearly
that at high angular momenta, the fully polarized case has a
low-energy state only at every 5th and 7th angular momentum,
respectively. In the case of six electrons the period of 5 becomes
clear only at higher angular momenta than those shown in Fig.~\ref{spec6},
due to the competition of two possible classical configurations of the
electrons in the Wigner molecule\cite{manninen2001}.

For polarized (spinless) electrons the tendency for localization is 
rather insensitive to the interparticle interaction.
It appears for long-range Coulomb interactions, as well as for short
range Gaussian interactions. Moreover, similar localization patterns 
occur for fermions as well as bosons~\cite{reimann2006}. In all these cases, 
the effect of the localization is seen as periodic oscillations 
in the energy as a function of the angular momentum.
There is one exception, however. The exactly solvable 
model of harmonic interparticle interaction does not 
show this periodicity\cite{johnson1993,ruuska2005}.
The reason is the large degeneracy of the energy states and
the fact that classically, all particles interacting by a 
repulsive harmonic potential are localized at the origin 
(otherwise, they would not be confined).

\begin{figure}[h]
\includegraphics[angle=-90,width=\columnwidth]{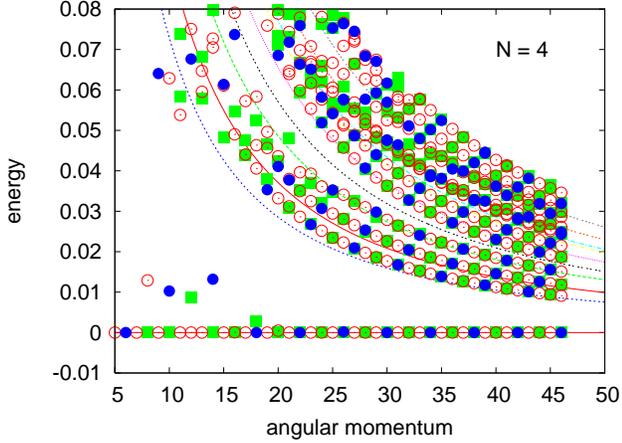}
\caption{Excitation energies as a function of the angular momentum.
Points are results of the exact diagonalization and the lines
from the model of classical vibrations.
Blue points: $S=2$, open circles: $S=1$, green squares: $S=0$.
}
\label{class4}
\end{figure}

Nikkarila and Manninen\cite{nikkarila2006} have shown that 
at large angular momenta the whole many-particle spectrum of
rotating polarized electrons can be quantitatively 
described by quantizing the classically determined vibration 
frequencies of the Wigner molecule. This suggests that the
charge excitations (vibrational modes) and the spin excitations
will separate in a similar fashion than in one-dimensional quantum
rings\cite{koskinen2001,viefers2004}, in which case the system
can be described by the Hamiltonian of Eq.~(\ref{mh}).

We determined the vibrational modes of classical electrons 
in a 2D harmonic confinement using a rotating frame, 
as in Ref.~\onlinecite{nikkarila2006}, and compared the results with
those obtained from the full quantum-mechanical CI calculation.

For $N=4$ electrons, the result  is shown in Fig. \ref{class4}.
The figure shows the excitation energies for each angular momentum,
i.e. the total energy of the excited state minus the lowest energy 
for the same angular momentum. The quantum-mechanical results
approach the classically determined energies when the angular
momentum increases. Different spin states of the QM calculation
are marked by different symbols. We can see clearly that the lowest
energy state and each vibrational state has a periodic pattern of spin
states.

\begin{figure}[h]
\includegraphics[angle=-90,width=\columnwidth]{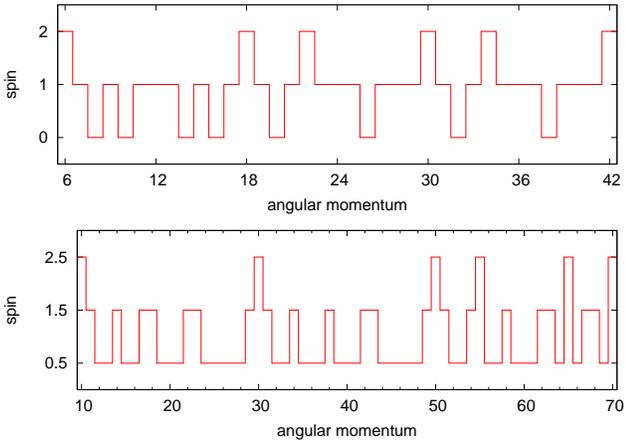}
\caption{Total spin of the lowest energy state as 
a function of the total angular momentum for $N=4$ and $N=5$.
}
\label{spin4and5}
\end{figure}  

The periodic appearance of the different spin states at
each vibrational mode in Fig. \ref{class4} gives further 
support to the separation of the spin and charge excitations.
Assuming the model Hamiltonian of Eq.~(\ref{mh}) one can use
group theory\cite{tinkham1964,koskinen2001,koskinen2002} to analyze
which spin state corresponds to which angular momentum. 
The Heisenberg coupling constant $J$ determines the energy 
splitting of different spin states, and its sign determines if 
the electron system is ferromagnetic or anti-ferromagnetic.
Indeed, the obtained spin states agree well with the prediction of the 
model Hamiltonian. However, the simple model can not explain 
quantitatively the energy differences between the different spin 
states. For example, for angular momenta 18, 22, 24, etc.,
the purely rotational band has spin states $S=2$ and $S=0$.
If the Heisenberg coupling would be ferromagnetic, the
$S=2$ state would have lower energy than the $S=0$ state, while
for anti-ferromagnetic coupling, this would be the other way around. 

Figure \ref{spin4and5} shows the total spin of the 
lowest energy states for 4 and 5 electrons as a function of the 
angular momentum. Clearly, we can see that neither of these cases 
show a simple period of $N$, as it would be expected from 
a simple Heisenberg model of localized electrons.

In the case of one electron per lattice site 
the Hubbard model approaches to the anti-ferromagnetic
Heisenberg model\cite{vollhardt1994}. Since here 
the electrons are localized in a Wigner lattice,
one should expect that the same model works equally well
(although now the
electrons are localized not by external potential, but by the mean
field created by the other electrons). For 
quasi-one-dimensional quantum rings this indeed was shown to be the 
case~\cite{koskinen2001,viefers2004},
as well as for electrons localized in the corners of triangular or
square quantum dots~\cite{jefferson1996}. 
However, the result of Fig. \ref{spin4and5} shows clearly that this
does not hold for electrons localized in a two-dimensional
harmonic confinement. 
\begin{figure}[h]
\includegraphics[angle=-90,width=\columnwidth]{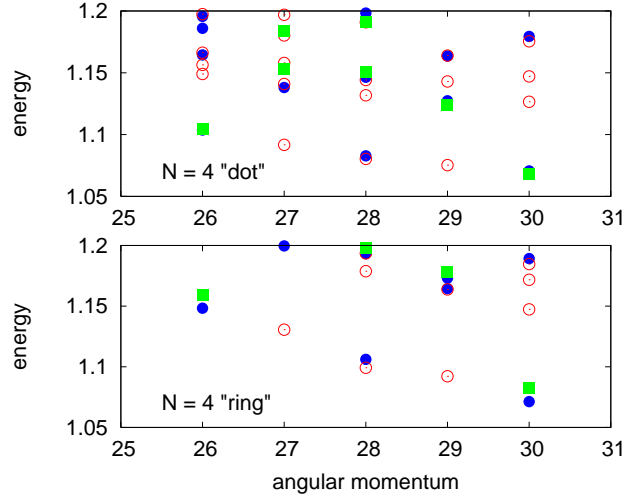}
\caption{Energy spectrum for four electrons in a quantum dot 
($\ell_{\rm min}=0$, $\ell_{\rm max}=11$)
compared to that of a quasi-one-dimensional quantum ring
($\ell_{\rm min}=3$, $\ell_{\rm max}=11$)
determined by restricting the basis as explained in the text. 
Blue points: $S=2$, open circles: $S=1$, green squares: $S=0$.
}
\label{ringdot}
\end{figure}  

We can mimic a quantum ring by restricting the 
single-particle basis from low and high angular momenta, i.e.
$\ell_{\rm min}\le \ell\le \ell_{\rm max}$.
Figure \ref{ringdot} shows part of the energy spectrum computed with 
this restricted basis compared to that of the full calculation.
The result for the 'ring' shows an anti-ferromagnetic 
spin arrangement in agreement with the Heisenberg
model\cite{viefers2004},
while for the dot the result is different. Note, however,
that in both cases we observe the separation of the lowest, purely 
rotational band, with the same sequence of spin states at each 
angular momentum.

\begin{figure}[h]
\includegraphics[angle=-90,width=\columnwidth]{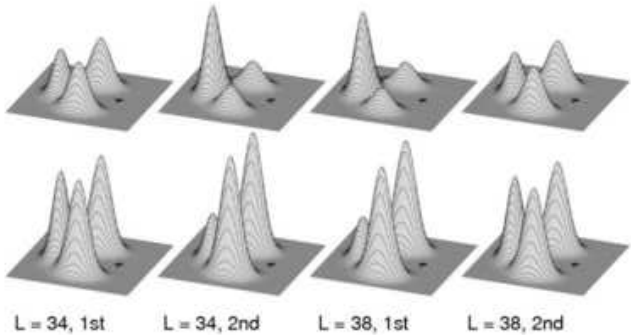}
\caption{Pair correlation functions for $N=4$ at angular
momenta $L=34$ and $L=38$ for the two lowest energy states.
The upper row shows the up-up correlation,  $g_{\uparrow, \uparrow}$, 
and the lower row the up-down correlation, $g_{\uparrow, \downarrow}$.
The position of the reference electron is shown as a cross.
}
\label{pc4}
\end{figure}  

Figure \ref{pc4} shows examples of the pair correlation functions,
Eq.~(\ref{pairc}), for four electrons in the purely rotational states.
We can see that for the fully polarized states, i.e. the $1$st 
state for $L=34$ and 2nd state
for $L=38$, the up-up and up-down pair correlations are similar due
to the symmetry of the ferromagnetic state (we show the $S_z=0$
result), while for spin singlet states ($S=0$) the pair correlations 
show a tendency for antiferromagnetism. It is important to note also that
the two states for $L=34$ and $L=38$ are similar,  but of opposite order
in energy. However, the energy differences are extremely small
as seen in Fig. \ref{ringdot}.

\begin{figure}[h]
\includegraphics[angle=-90,width=\columnwidth]{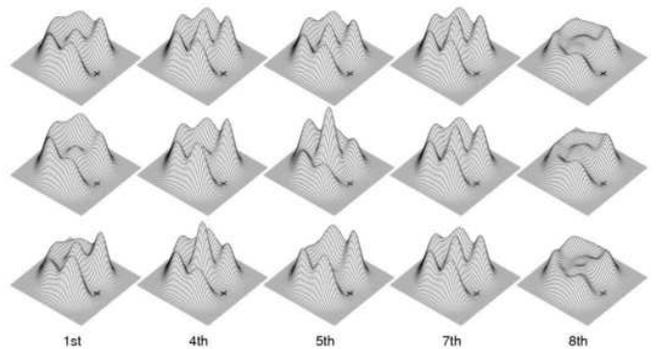}
\caption{Pair correlation functions for $N=7$ at angular
momenta $L=51$ for the lowest energy state, and for
the 4th, 5th, 8th and 9th excites states.
The uppermost row shows the total correlation function,
$g_{\uparrow, \uparrow} + g_{\uparrow, \downarrow}$, 
the center row the up-up correlation $g_{\uparrow, \uparrow}$, 
and the 
the lowest row the up-down correlation, $g_{\uparrow, \downarrow}$. 
The position of the reference electron is shown as a cross.
}
\label{pc7}
\end{figure}  

For polarized electrons it is possible to perform an exact diagonalization 
for large angular momenta where the localization of electrons is 
clearly seen\cite{reimann2006}. The spin degree of freedom, however,
increases the available Fock space drastically. For example, for
$N=7$ or 8 electrons  we are limited to 
study only the region up to $\nu\approx 1/3$. However, we can see that the
localization begins already in this region. Figure \ref{pc7}
shows the pair correlation function for 7 electrons at $L=51$
corresponding to the filling factor $\nu=2/5$. The total correlation
function, Eq.~(\ref{pctotal}), shown in the upper row of Fig. \ref{pc7},
shows clearly that the electrons begin to localize in a geometry where
one electron is at the center, and the remaining six electrons form 
a hexagon around it. The lowest-energy state at this angular momentum
is well separated from the excited states and 
can be approximatively described by the Halperin-Haldane wave
function as mentioned above. It is interesting to note, that in this state
the electrons are not as well localized as in the excited states
(4th, 5th and 7th state).
The 8th state is above the large energy
gap seen in Figs.~\ref{spec7} and~\ref{spec7d}. 
In this state, the electrons are not 
as clearly localized as in the lower-lying states.  
This is in agreement with the beginning
separation of the spin excitations from the charge excitations:
Below the large energy gap in Fig.~\ref{spec7d} all the excitations 
are spin excitations, which do not markedly change the pair correlation
function, while above the gap, the excitations include a charge
excitation which necessarily changes the pair correlation.
We checked the pair correlations for all the states 
from $L=51$ to $L=75$. In all cases, the states 
below the energy gap are similar, showing the localization 
of electrons which gradually becomes stronger when the 
angular momentum increases~\cite{reimann2006}.

Figure~\ref{pc7} shows also the up-up and up-down 
pair correlations. These support the view that the
low-energy excitations are mainly spin excitations.
The three $S=1/2$ states below the energy gap 
(i.e. the 1st, 4th and 5th state)
show different up-up and up-down correlations, while the 
total correlation is nearly similar.
In the case of the lowest energy state, the
up-up and up-down correlations are in agreement with
the Halperin-Haldane state: The repulsion between
opposite spins is larger than that between the same spins.

The pair correlation functions shown in Fig.~\ref{pc7}
do not show much similarity to those we obtained in the  
anti-ferromagnetic Heisenberg model.

Note, that in the fully polarized  case, $S=7/2$,
the up-up and up-down correlations, calculated for $S_z=1/2$, 
are identical, as expected.

\section{Conclusions}

We have studied rotational states of interacting electrons in a 
two-dimensional quantum dot confined by a circular
harmonic potential. The many-particle Hamiltonian was solved
exactly by numerical diagonalization in a basis 
restricted to the lowest Landau level. The results were 
analyzed in terms of the total angular momentum and spin.
Magnetic fields were not considered explicitely, and consequently,
no Zeeman splitting was introduced.

The smallest possible angular momentum in the LLL 
is a 'double maximum density droplet' with 
$L=N(N-2)/4$ corresponding to
the filling factor $\nu=2$. Increasing the angular momentum
the total spin of the system increases oscillating
between zero and the next maximum until it reaches the
fully polarized ferromagnetic state of the maximum 
density droplet. This behavior was found to be the same
for the long range Coulomb interaction and for a contact 
interaction.

Beyond the MDD, the lowest energy states are spin-waves 
of the ferromagnetic state in agreement with previous 
calculations with different periodic boundary conditions.
Increasing the angular momentum further, the lowest-energy
states seem to consist of ferromagnetic domains, while the
total spin of the system is zero.

For low total spin we did not find any vortices as
low energy excitations, although for the polarized case they 
appear as the lowest energy states at certain angular momenta.

The excitation spectrum shows a clear reflection symmetry
around angular momentum $L=N(N-1)$ corresponding to a filling
factor $\nu=1/2$. In the neighborhood of angular momenta 
$L=N(N-1)/2+N^2/4$ and $L=3N(n-1)/2-n^2/4$, corresponding to
filling factors 2/3 and 2/5 respectively, the spectra look 
very similar. Both states have $S=0$ and a large energy gap to the
first excited state. In fact, these energy gaps 
are the largest in the whole
region from $\nu=1$ to $\nu=1/3$, apart from those in 
the immediate vicinity of filling factor $\nu=1$.
The lowest-energy state with filling factor $\nu=2/5$ can be
rather accurately approximated by the Halperin-Haldane
generalization of the Laughlin wave function, at least
for small numbers of electrons.

In addition, the spectra are symmetric around 
angular momenta $L=N(N-1)/2$ and $L=3N(N-1)2$ corresponding
to filling factors 1 and 1/3, respectively. 
Especially, the states around $\nu=1/3$
seem to have similar spin wave excitations as the states 
around filling factor one. 
Very recently, 
Dethlefsen {\it et al.}\cite{dethlefsen2006}
have studied the details of the excitation in the region of
$\nu\approx 1/3$ with comparison to experiments. 

At large angular momenta, the electrons start to localize.
In the case of four electrons, accurate results  could be obtained
up to angular momenta $L\le 46$, corresponding to a filling factor
smaller than $\nu=1/7$. In this case the results
showed a clear separation of charge-like and spin-like 
excitations. The charge excitations could be quantitatively
explained by quantization of the classical vibrational 
modes of the localized electrons. However,
contrary what we expected (for localized electrons)
the spin-excitations could not 
be explained within a simple Heisenberg model.

{\bf Acknowledgments:}
We thank B. Mottelson, P. Hawrylak, J. Jain, S. Viefers, K. Vyborny
and Y. Yu for discussions. 
This research was financed by the Academy of Finland, 
the Swedish Research Council, the Swedish Foundation for Strategic Research, 
and NordForsk.

\end{document}